\documentstyle[url,pre,aps,multicol,epsf,epsfig,amsmath,epic,eepic,rotate]{revtex}
\begin{document}
\draft
\title{Attractive forces between anisotropic inclusions in the membrane of a vesicle}
\author{R. Holzl\"ohner
\thanks{Current address: Dept.~of Electrical Eng.~, TRC 201 B, University of Maryland
    Baltimore County, 1000 Hilltop Circle, Baltimore MD 21250, USA. \
    e-mail: \tt holzloehner@umbc.edu}}
\address{
 Institut f\"ur Theoretische Physik,
 Technische Universit\"at Berlin, Hardenbergstra{\ss}e 36, 
 D-10623 Berlin, Germany}

\author{M. Schoen
\thanks{e-mail: \tt schoen@wpta9.physik.uni-wuppertal.de}}
\address{
 Fachbereich Physik --- Theoretische Physik,
 Bergische Universit\"at Wuppertal, Gau{\ss}stra{\ss}e 20,
 D-42097 Wuppertal, Germany}
\date{April 20, 1999}

\maketitle

\begin{abstract}
  The fluctuation-induced interaction between two rod-like, rigid inclusions
  in a fluid vesicle is studied by means of canonical ensemble Monte Carlo simulations.
  The vesicle membrane is represented by a triangulated network of hard spheres.
  Five rigidly connected hard spheres form rod-like inclusions that can leap
  between sites of the triangular network. Their effective interaction potential
  is computed as a function of mutual distance and angle of the inclusions. 
  On account of the hard-core potential among these, the nature of the potential 
  is purely entropic. Special precaution is taken to reduce lattice artifacts
  and the influence of finite-size effects due to the spherical geometry. Our
  results show that the effective potential is attractive and short-range 
  compared with the rod length $L$. Its well depth is of the order of 
  $\kappa/10$, where $\kappa$ is the bending modulus.
\end{abstract}

\pacs{PACS: 87.22.Bt,68.35.Md,02.70.Lq}

\begin{multicols}{2}

\narrowtext

\newcommand{\kBT}{{k_{\rm B} T}}
\newcommand{\beq}{\begin{eqnarray}}
\newcommand{\eeq}{\end{eqnarray}}
\newcommand{\non}{\nonumber}
\newcommand{\msf}{\mathsf}
\newcommand{\ovl}[1]{\overline{#1}}
\newcommand{\Ronfigure}[4]{
  \begin{figure}[tbp]
    \begin{center}
    \scalebox{#2}{\includegraphics{#1}}
    \end{center}
    \caption{#3}
  \label{#4}
\end{figure}
}
\newcommand{\Latexfigure}[3]{%
\begin{figure}[tbp]%
  \center{  \input{ #1}
\caption{#2} \label{#3} }
\end{figure}
}
\newcommand{\LatexfigureTWO}[4]{%
\begin{figure}[p]%
\begin{center}
 \begin{minipage}{\columnwidth}
 \input{#1}
 \vspace{-2cm}
 \input{#2}
 \caption{  {#3} \label{#4} }
 \end{minipage}
\end{center}
\end{figure}
}
\section{Introduction}
\label{sec:intro}

 Lipid membranes are interesting systems in statistical physics and
are the subject of many theoretical and experimental investigations
because of the abundance of effects they exhibit.
Often {\em fluid} membranes are considered, which feature the unusual
combination of finite bending stiffness and vanishing in-plane 
shear stress. This is caused by the constituent lipid molecules
which can be sheared against each other, but resist bending normal to 
the plane \cite{Lip98b}.

Biological membranes also contain inclusions, i.~e.~impurities that
differ from lipid molecules chemically and mechanically.
Inclusions are embedded in the membrane and
can diffuse laterally. Examples range from rather large
inclusions such as proteins 
and polymers to very small bodies such as so-called {\em gemini}
comprising two lipid molecules whose head groups are chemically 
bonded \cite{ACK92}.
In general, any membrane component that deviates in its mechanical 
properties from lipid molecules will be called inclusion. Most of these
are rather rigid and thus, the presence of an inclusion
locally stiffens the ambient membrane.
Inclusions diffuse within the membrane with typical speeds of a 
few microns per second \cite{ACK92}.\\

 Forces between membrane inclusions currently receive considerable 
interest \cite{Fou96,Gou93,Par96,Gol96b,Net97b,Wei98}.
They can be divided into two classes, {\em direct} forces due to
electrostatic and van der Waals interactions, and {\em indirect}
forces which are mediated by membrane fluctuations. The latter are
of interest here. However, indirect fluctuation forces between 
membrane inclusions
should not be confused with depletion forces \cite{Yam98a,Goem98}
(although both are entropic in origin) that
exist when small particles are depleted from the gap between bigger
ones. Indirect inclusion interactions were investigated theoretically, 
both for isotropic (rotationally invariant in the membrane plane) 
\cite{Fou96,Gou93,Par96,Net95a,Dom98},
anisotropic (e.~g.~rod-like) inclusions \cite{Par96,Gol96b},
and under lateral membrane tension \cite{Wei98}.
The above quoted contributions focus on the range $\xi \gg s \gg L$,
where $s$ is the center-center distance between two inclusions whose 
linear, in-plane size is $L$, see Figure \ref{inclusions}, and $\xi$ is the
persistence length, that is the distance in the membrane over which the 
correlation of the surface normals decays \cite{Gou93}.
Based upon perturbative approaches, it was found
that there is an (attractive) long-range interaction potential of the form
\hbox{$\Phi \propto -\kBT\;(L/s)^4$}, both for isotropic \cite{Gou93} and 
anisotropic \cite{Par96,Gol96b} inclusions ($k_{\rm B}$ and $T$ are
Boltzmann's constant and temperature, respectively). 
However, its magnitude was
predicted to be much smaller than $\kBT$ over the range of $s \gg L$.

Short range attractive interactions ($s < L$) have been predicted by Netz
\cite{Net97b}
who treated the interactions between stiff inclusions in a membrane
analytically. For his model Netz obtains a logarithmically decaying,
attractive interaction potential whose magnitude increases with increasing
membrane stiffness (see Sec.~\ref{sec:The_Model},\ref{sec:Results}). 
Other studies dealing with short-range
interactions between membrane inclusions were performed by Dan
et al.~\cite{Dan94} and Aranda\--Espinoza et al.~\cite{Ara96}
who considered membrane
Hamiltonians with contributions from compression (expansion), spontaneous
curvature of the membrane, and bending stiffness. These calculations have
been carried out in the limit of vanishing temperature where membrane
fluctuations do no longer exist. Also, $s$ in
\cite{Dan94,Ara96} is typically of the order of the membrane thickness,
whereas the papers 
\cite{Gou93,Par96,Gol96b} assume a thickness much smaller than $s$.

The present paper is also concerned with short-range interactions 
but between rod-like inclusions for $s \approx L$
embedded in the surface of a vesicle.
Throughout this paper we are exclusively concerned with the 
case of zero spontaneous curvature.
To the best of our knowledge, numerical simulations 
for three-dimensional fluctuating membranes with finite 
bending stiffness and inclusions, which are considered here, 
have not yet been carried out.
\\

The remainder of this paper is organized as follows.
Model and simulation algorithm are detailed in Sec.~\ref{sec:The_Model}.
In Sec.~\ref{sec:Pair distribution functions}, pair distribution 
functions are introduced. 
Details of their numerical determination are presented in
Sec.~\ref{sec:Numerical_Details}.
Sec.~\ref{sec:Results} is devoted to a presentation 
of the results obtained in this work.
The paper concludes in Sec.~\ref{sec:Discussion and Conclusions}
with a summary and discussion.
\Ronfigure{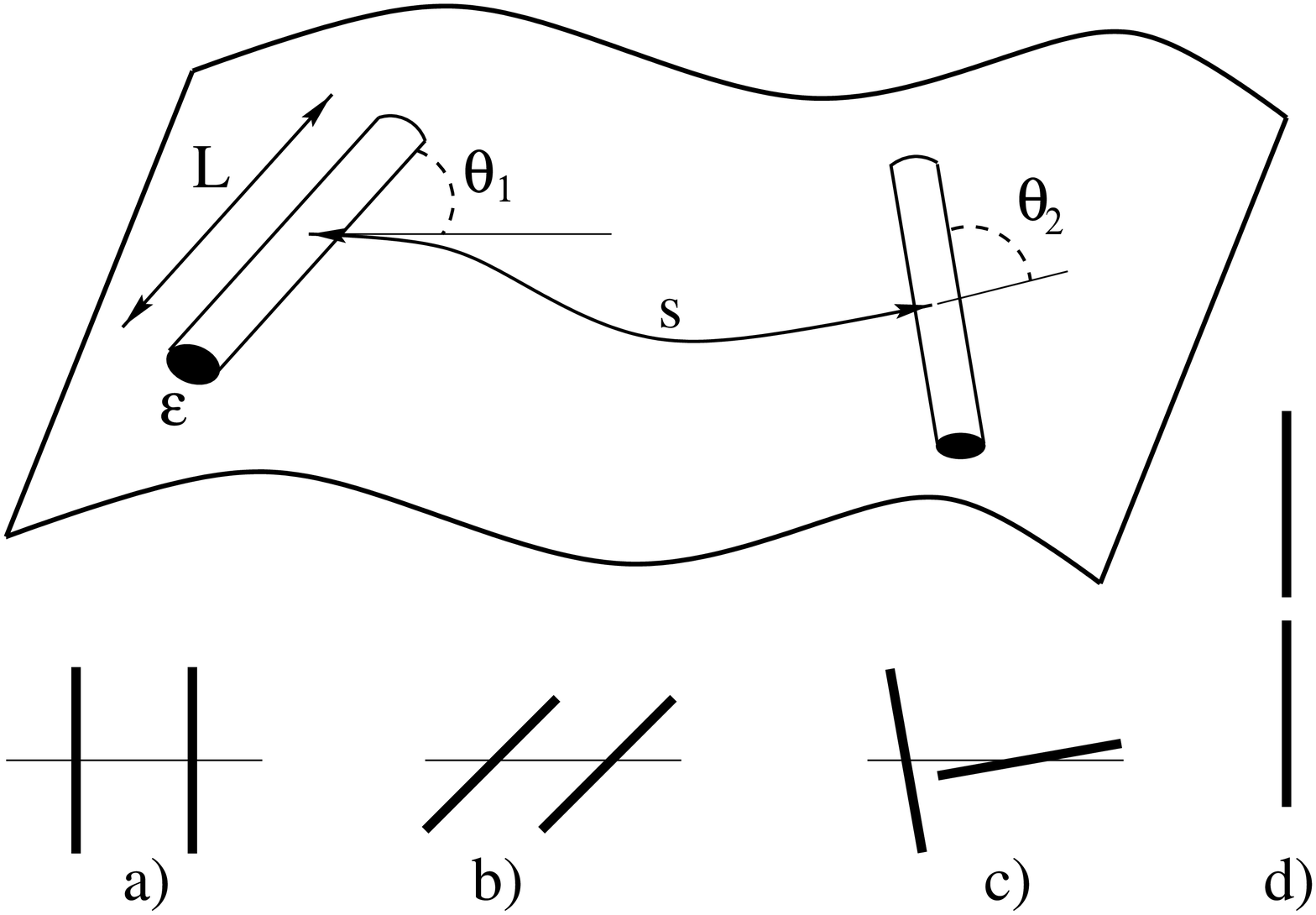}{0.3}{
  Curved membrane with
  two rod-like surface inclusions of length $L$ and width 
  $\epsilon \ll L$, separated by the center-to-center distance $s$.
  The rods are rotated by the in-plane angles $\theta_1,\theta_2$ 
  against the connection vector. Sketches a) and b) show two rods in 
  parallel settings. A perpendicular arrangement similar to 
  c) will be called ``T-formation'',
  a setting such as d) ``in-line formation''.
}{inclusions}
\section{The Model}
\label{sec:The_Model}
The approximations employed in \cite{Gou93,Par96,Gol96b}
are only valid in the range $s \gg L$ and break down for
the interesting case of $s \approx L$.
Also, the membrane can no longer be regarded as a continuous surface, 
as the discrete lipid network becomes more and more influential. Hence,
numerical simulations are required to obtain quantitative results.
In this regard the Monte Carlo method provides a particularly 
powerful technique by which thermophysical properties of equilibrium
systems can be computed in a rather simple and straightforward manner
\cite{SCH99}.

The model membrane consists of $N$ hard spheres
of diameter $a$, connected by rigid bonds (tethers) of length 
$r$ with \hbox{$a \leq r \leq \sqrt{3} a$}.
This model has been employed previously
\cite{Gom97a,Gom96,Ips95,Kan87a}. Membrane fluidity is realized
by the {\em bond-flip} algorithm first proposed in \cite{Kaz85}.
On a closed triangular network, each bond can be regarded as one of the 
diagonals in the quadrilateral formed by the four surrounding bonds. 
The bond-flip algorithm rotates the bond within this 
quadrilateral, so that it represents the other diagonal after the operation.
This method allows for vertex diffusion, as any triangulation 
can be transformed into any other \cite{Kaz85}. 
The bending energy is computed from the Helfrich Hamiltonian
\cite{Hel73} by integrating over the surface $S$
\begin{equation} \label{Helfrich_energy}
  {\cal H} =  \int \!{\rm d}S
    \left(\frac{\kappa}{2} H^2 + \bar\kappa K \right) \;,
\end{equation}
where $H \equiv c_1+c_2$ is the sum of the principal curvatures on the 
surface, $K \equiv c_1\:c_2$ the Gaussian curvature, and $\kappa, \bar\kappa$
are, respectively, the bending modulus and the Gaussian modulus.
The term $\int \!{\rm d}S \; K$ is constant in fixed surface topology
due to the Gau{\ss}-Bonnet theorem \cite{Hel73}.
The term $\int\!{\rm d}S\;H^2$ is discretized as in \cite{Gom96},
\newcommand{\tat}{ {\bf n}_{\triangle_a} }
\newcommand{\tbt}{ {\bf n}_{\triangle_b} }
\newcommand{\tattbt}{ \triangle_a,\triangle_b }
\newcommand{\avg}[1]{ \left\langle {#1} \right\rangle  }
\begin{equation} \label{triangle_sum}
  \frac{\kappa}{2} \int\!{\rm d}S\;H^2 \approx
	4\pi\,\kappa + 
	\sqrt{3}\,\kappa \sum_{i=1}^{N_B} \bigl(1-\tat \tbt\, \bigr) \;,
\end{equation}
where the sum runs over all $N_B = 3(N-2)$ 
bonds in the network. Each bond has two
adjacent triangles $\triangle_a(i)$ and $\triangle_b(i)$, 
whose outer unit normals are denoted by $\tat$ and $\tbt$.

\Ronfigure{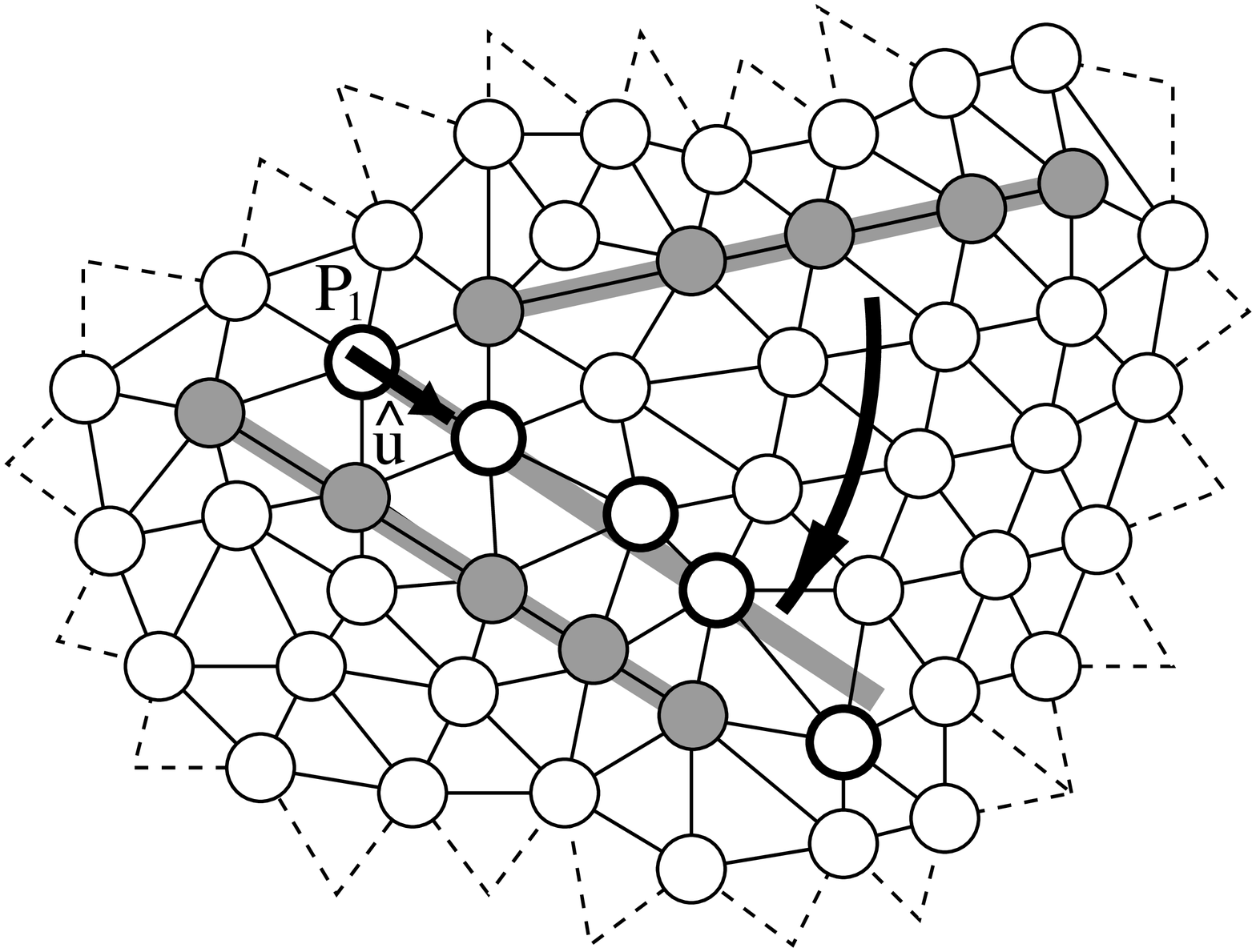}{0.3}{
  Portion of a triangular network of hard spheres with two
  inclusions, consisting of five vertices each.
  To the ambient membrane, vertices and bonds within 
  inclusions appear ``frozen''
  and cannot be moved (or flipped) by an MC-step. 
  An inclusion can leap as an entity (arrow)
  by releasing the old vertices and freezing a new set 
  (bold circles), after having moved the vertices.
  ${\bf \widetilde P}_1$ is the position of the selected
  vertex from which the new rod is tried to be placed
  along the tangential direction ${\bf \hat u}$, see text.
}{netz}
 The simulation algorithm consists of two independent parts.
 The first one follows the one described in \cite{Gom96}.
One Monte Carlo step (MCS) consists of attempting to
move $N$ randomly selected vertices to new positions within a 
cube $[-\rho,\rho]^3$, centered at their current positions.
Next, $N$ bonds are selected and attempted to be flipped.
Both processes are accepted or rejected on the basis of their
associated change in energy according to the Metropolis algorithm
\cite{ALL87}.

 In addition, mobile, rod-like inclusions are embedded in the
membrane. They consist of five rigidly connected hard spheres.
None of the spheres in this ``frozen'' rod-like configuration
can be moved independently, nor can the four inner
bonds be flipped. A rod and its ambient membrane are thus 
similar to a spine in which some
of the vertebrates are fused. Their biological equivalent would come closest
to (multi-)gemini, 
since the inclusion constituents are equal to those of the plain membrane.

 The inclusions can move as an entity, thereby resembling lateral diffusion.
This is accomplished by the ``rod-leap'' algorithm, described below,
which is carried out in the second step of the simulation.
A new set of five vertices is selected, which may overlap with
the old one. To find the new vertices, a random vertex next to
the old rod is selected and the four remaining vertices are chosen
along the direction of a random unit tangential vector ${\bf \hat u}$.
Subsequent vertices in the rod must be connected by a bond,
as shown in Figure \ref{netz}.
Then a Monte Carlo move is attempted to simultaneously move 
the last four of the new vertices to the new positions
\begin{equation}
  {\bf \widetilde P}_n = 
    {\bf \widetilde P}_1 + ( {\bf \hat u} \cdot {\bf P}_n ) \, {\bf \hat u} \;,
\end{equation}
where ${\bf \widetilde P}_1 = {\bf P}_1$ is the position of the selected
vertex and the ${\bf P}_n$, $(n=1 \ldots 5)$
are the old vertex positions. 
The orientation of the 
local tangential plane next to ${\bf P}_1$ is found by averaging all the 
neighboring triangle normals.

The effect of placing a rod somewhere is to flatten the membrane 
locally. 
Note that this does not affect the surface topology. No holes or
contact angles at the membrane-inclusion boundary are introduced
and hence the Gau{\ss}-Bonnet theorem remains valid.
If finite contact angles are assumed \cite{Wei98,Dom98}, the
number of the inclusions must be kept constant, whereas
in the present case the Gau{\ss}-Bonnet theorem would 
remain valid even if the number of inclusions were varied.
This flattening effect quenches fluctuations normal to surface of 
the vesicle.
On the other hand, inclusions hardly increase the net bending energy
of the ground state vesicle (sphere) of \hbox{$E_0 = 8 \pi\kappa$}
\cite{Gom96}. 
Figure \ref{SnapShot} shows a snapshot of a simulation with 
$N=1012$ vertices.

The total number of vesicle vertices $N$ and number of rods $N_R$ must
be carefully chosen to ensure that the vesicle is not too strongly 
perturbed by the rods. Also, the impact of multi-body effects among 
the rods is crucial for the reliability of the simulations.
The rods on the membrane can be considered 
a lattice gas that should be as dilute as possible.
In the present simulation, $N_R=4$ rods were placed on a vesicle of
$N=1012$ vertices. With this choice, only $2\%$ of the vertices are 
occupied by inclusions. The average
rod length of $L=4\avg{r}$, 
where $\avg{r} \approx a(1+\sqrt{3})/2$ is the mean bond length,
is only about half as long as the average vesicle radius of gyration 
that is computed to be $\avg{R_g}=8.13\avg{r}$ for $\kappa/\kBT = 7.5$. 
An alternative measure of coverage of the membrane by inclusions is obtained 
by assigning a disk of size 
$A_L = \pi\, (L/2)^2$ to each rod. The total area covered by the disks
\hbox{$N_R A_L/\avg{A}$} is about $6\%$ 
of the average vesicle surface area ($\avg{A} \approx 841 \avg{r}^2$).
Thus, we conclude that with the above choice of constants,
the vesicle is not strongly perturbed by the inclusions and many-body
effects are expected to be negligible (see also below).
\Ronfigure{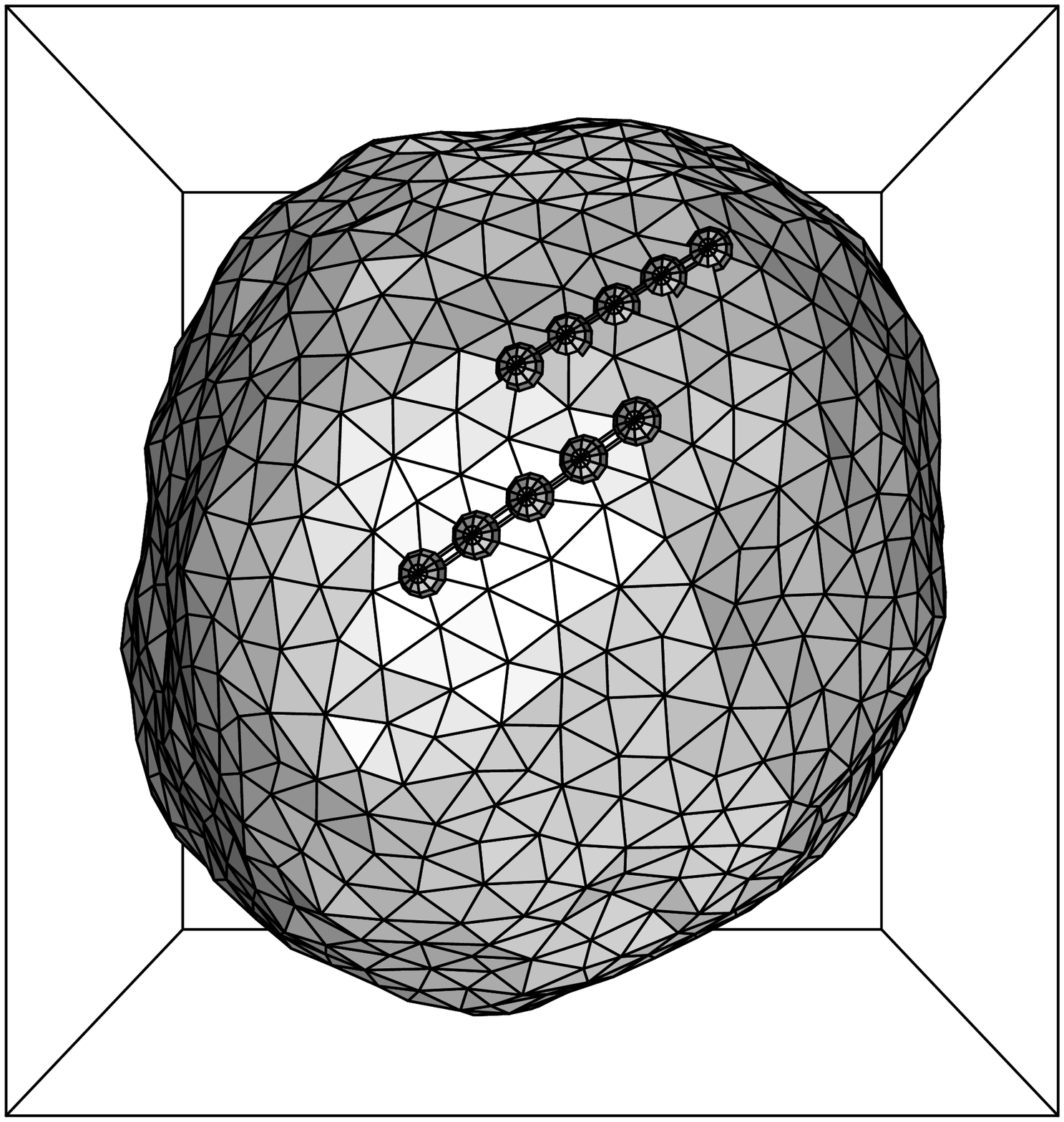}{0.42}{
  Simulation snapshot of a vesicle with $N=1012$ vertices and
  $\kappa/\kBT=7.5$.
  The shades on the surface may be thought of as reflections from
  a spotlight. Two inclusions with their incorporated
  spheres are visible on the membrane.
}{SnapShot}

Thermal fluctuations induce a finite effective surface tension $\tau$ 
in fluid membranes. Therefore, in principle $\tau$
introduces an additional length scale. However, this tension was estimated
as $\tau \approx \kBT \xi^{-2}$ \cite{Dav91} and thus is very small 
in the regime that is of interest here. In the case of biological 
membranes, experiments show \cite{Bro76} that the effective surface tension 
is negligibly small. 
\section{Pair distribution functions}
\label{sec:Pair distribution functions}
The goal of the present work is the computation of the fluctuation-induced 
interaction potential of the inclusions 
$\Phi_{\rm RR}(s,\theta_1, \theta_2, \kappa)$.
The total energy of a given triangulation ${\cal T}_N$, as given by 
\eqref{Helfrich_energy} and \eqref{triangle_sum}, depends on the
positions of the vertices ${\bf X}_1\dots{\bf X}_N$ 
and on the $N\times N$
connectivity (adjacency) matrix $\mathsf S$, where ${\mathsf S}_{ij}=1$ if 
vertices $i$ and $j$ are connected by a bond and zero otherwise.
The partition function can then be written as \cite{Ips95}
\beq \label{split-up_part_sum}
  {\cal Z} &=& 
  \int\! {\rm d}{{\cal T}_N}\;
   {\exp\bigl(-\beta ({\cal H}[{\cal T}_N] +  U_{\rm bond}) \bigr)} \non\\
   &=& \sum_{\msf S} 
   \int\! \Bigl(\prod_{i=1}^N {\rm d}{\bf X}_i\Bigr)\:
     \exp\bigl(-\beta ({\cal H}[{\cal T}_N] +  U_{\rm bond}) \bigr) \;.
\eeq
Here, $U_{\rm bond}$ is introduced as a constraint imposed on the bonds 
between neighboring hard spheres forming the membrane. It is zero if all
bond lengths $r_i$ are in the range 
\hbox{$a \leq r_i \leq \sqrt{3} a$}
and infinite otherwise. In our model, the only effect of adding inclusions
to the membrane is to prohibit certain triangulations that do not adhere
to the conditions described in the preceding section.
In \eqref{split-up_part_sum}, this can be accounted for by
extending the definition of $U_{\rm bond}$ so that it diverges also
if ${\cal T}_N$ cannot accommodate the $N_R$ inclusions.

 Triangulations can be transformed into each other by means
of the bond-flip algorithm and thus no vertex is different from 
the others. Consequently, the average energy of a triangulation with, say,
two inclusions only depends on their mutual distance $s$ and
angles $\theta_1$, $\theta_2$, see Figure \ref{inclusions}.
This is the basic assumption of the present work. 
Strictly speaking, the distance $s$ is the length of the shortest 
geodesic on the surface that connects the centers of the two inclusions.
For $s \ll \xi$, however, the Euclidean distance can be taken.
The inclusion interaction is purely entropic and not a result of direct
molecular interaction. However, an effective interaction potential
$\Phi_{\rm RR}(s,\theta_1,\theta_2,\kappa)$ can be defined as a 
potential of mean force (PMF) \cite{HAN86} by the relation
\begin{equation} \label{PMF}
  \Phi_{\rm RR} + \Phi_{\rm ma} = -\kBT\, \ln g^{(2)}\;,
\end{equation}
where $\Phi_{\rm ma}$ is the mutual
avoidance potential of the rods and 
$g^{(2)} = g^{(2)}\bigl(s,\theta_1,\theta_2,\kappa\bigr)$ 
is the pair distribution function of the inclusions on the 
triangulation.\\

The potential $\Phi_{\rm RR}$ was calculated analytically in 
\cite{Gol96b} for the range $\xi \gg s \gg L$. Surprisingly, 
the result depends on the {\em sum} of the angles
$\theta_1+\theta_2$ between the rods and the connecting vector.
However, $\theta_1+\theta_2$ is the same for the T-formation 
(Figure \ref{inclusions} c) and, for example, the case
\hbox{$\theta_1=\theta_2=45^{\circ}$} (Figure \ref{inclusions} b). 
For small distances $s < L$ that are mainly considered in this article,
the latter case is essentially a parallel side-by-side
position, and thus parallel and
perpendicular relative orientations (Figure \ref{inclusions} a,c)
would be indistinguishable.
For this reason, the angle {\em difference}
\hbox{$\theta := |\theta_1-\theta_2|$} is
a better variable here for $s \approx L$. On the other hand, one
has to take into account that for $s \geq L$, the parallel side-by-side
position and e.~g.~the in-line positions (Figure \ref{inclusions} a,d)
are then indistinguishable.
\section{Numerical Details}\label{sec:Numerical_Details}
 We approximate $g^{(2)}$ by a histogram $g^{(2)}_{ij}$.
To do so, a rod pair histogram $n_{ij}$ of \hbox{$25 \times 25$} 
entries is generated.
After regular intervals during the Monte Carlo simulation, all
$N_R(N_R-1)/2$ pairs of the $N_R$ rods are considered. If
$s \leq s_{\rm max}$, where $s_{\rm max}$ is a cut-off
distance small compared to $\langle R_g \rangle$,
$n_{ij}$ is incremented by one.
Here, the indices
$i$ and $j$ are rounded to the closest integer
$i \simeq 25\; s/s_{\rm max}$ and 
$j \simeq 25\; \theta/(\pi/2)$ (rods are head-tail symmetrical).
For $s_{\rm max}$, a length of $1.5 L$ is used.
The relationship between $n_{ij}$ and $g^{(2)}_{ij}$ is
\begin{equation} \label{pair distribution_2}
  g^{(2)}_{ij} = 
    \frac{n_{ij}}{n_{\rm tot}}\:\frac{\avg{A}}{\avg{A_i}}\;,
\end{equation}
which is valid for $s \leq s_{\rm max}$. Here, 
$n_{\rm tot}$ is the
total number of pairs of rods considered
(some of which with $s > s_{\rm max}$ and thus 
\hbox{$n_{\rm tot} \geq \sum_{i,j} n_{ij}$}), 
and $\avg{A_i}$ is the average surface area of a strip 
on the vesicle with 
\hbox{$s \in [i,\: i+1] \!\times\! s_{\rm max}/25$},
analogous to the area between two parallels on the globe.
\Ronfigure{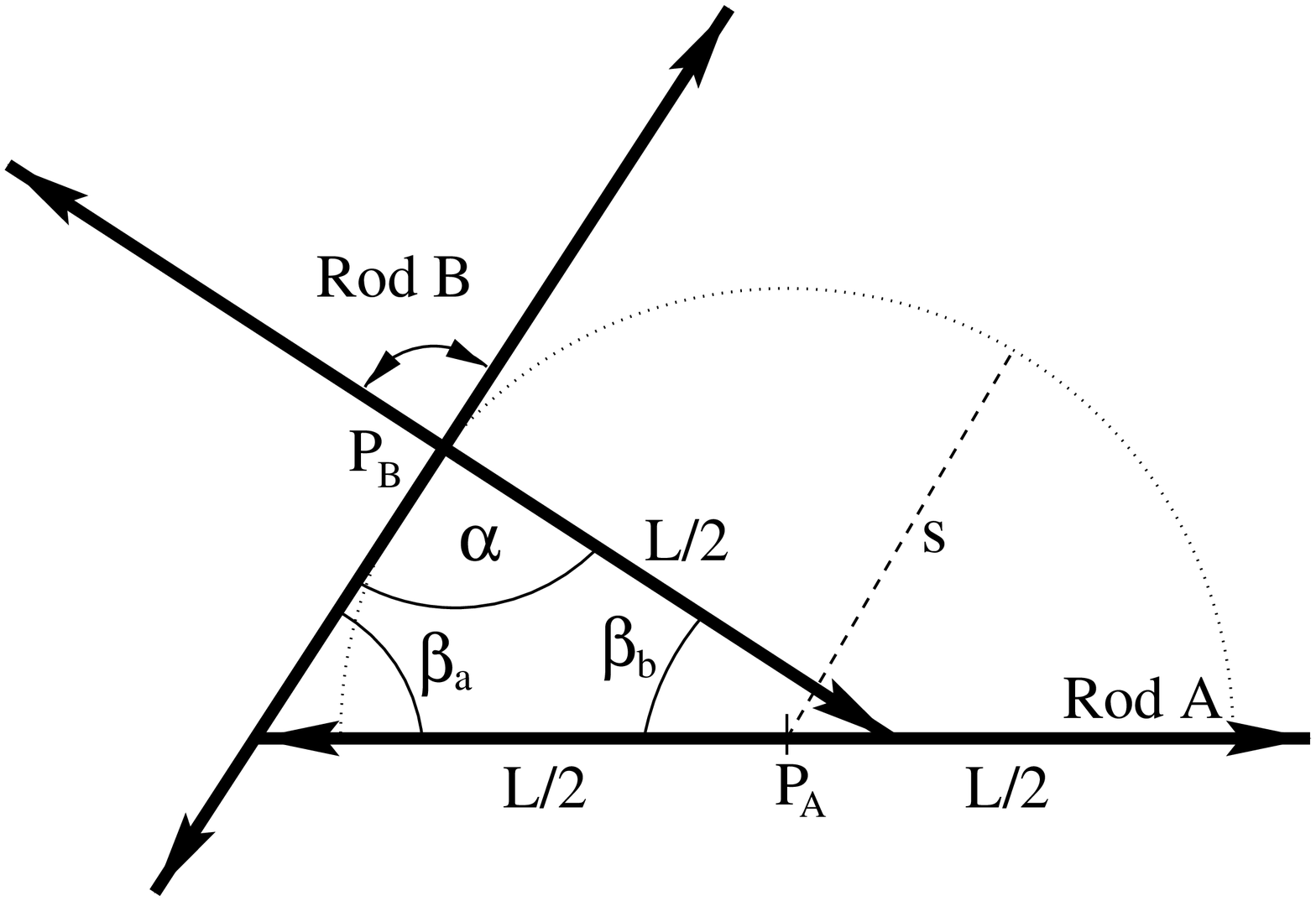}{0.3}{
  If two idealized rods $A$ and $B$ have a center-center distance 
  $\ovl{P_A\,P_B} = s$ of less than $L$, the usual accessible
  rotational angle of rod $B$ around its center $P_B$ of $\pi$ 
  is reduced to $\pi-\alpha = \beta_a+\beta_b$.
}{winkel}
The pair distribution function $g^{(2)}$ is unity for a uniform
inclusion distribution on an exactly spherical vesicle. However,\\
a)
the vesicle shape deviates from the sphere especially for 
small $\kappa$.\\
b) 
Since inclusions are bound to vertices, ``quantization'' effects 
in the density can be expected if $s$ becomes comparable to $\avg{r}$.\\
c)
In addition, a peak in $g^{(2)}$ for $s \approx a$ (the smallest
possible distance) can be expected for the following reason: 
An inclusion straightens a row of vertices, thereby forcing
its vicinity to remain closer to the perfect, flat and evenly spaced
lattice. Trial moves that consist in placing a rod parallel to an
existing one might be less likely to violate a bond length condition
because the lattice is more uniform there (closer to the regular
hexagonal lattice).\\
Effects b) and c) are typical lattice artifacts and therefore unphysical.
They are eliminated by the procedure outlined below.\\

 We are only interested in the variation in $g^{(2)}$
caused by the increased probability of placing inclusions close to
each other, for the membrane there is less corrugated normal 
to the plane than on average.
In order to remove artifacts a-c, normalization runs
are performed, according to the following prescription.
The rod-leap MC step conforms in all details to the prescription
described in the last section, except that before the bending energies of
the new and previous rod positions are compared, the new vertices are 
moved back to their original positions. In other words, in the normalization
run, each trial move is accepted if no bonds would be broken in moving
the vertices to the positions ${\bf \widetilde P}_n$,
but since vertices are never actually moved there, the lattice remains unchanged.
This yields a normalization histogram $\widetilde n_{ij}$ and analogous to
\eqref{pair distribution_2} a pair distribution function $\widetilde g^{(2)}$.
As a consequence, the moves in the normalization runs share all three types of 
artifacts a-c, but are insensitive to differences in bending energy.
This allows to compute $\Phi_{\rm RR}(s,\theta)$ by
\begin{equation} \label{potential}
  \frac{{\Phi_{\rm RR}}_{,\;ij}}{\kBT} \stackrel{~}{=}
    - \ln  \frac{\;g^{(2)\;}_{ij}}{\widetilde g^{(2)}_{ij}} =
    - \ln  \frac{n_{ij} \; \widetilde n_{\rm tot}}
      {\widetilde n_{ij} \; n_{\rm tot}} \;.
\end{equation}
Inclusions cannot overlap with others which implies a mutual 
avoidance condition of \hbox{$\sin\theta > 2s/L$} as shown in 
Figure \ref{winkel}. Also, the center-center distance cannot be smaller 
than a hard sphere diameter, $s \geq a$. The mutual avoidance potential
$\Phi_{\rm ma}$ from \eqref{PMF} is infinite in these 
regions and zero elsewhere.

Depletion effects \cite{Yam98a,Goem98} play no role in the present 
simulation, since the lipids (free vertices) as the ``small particles'' 
in the membrane network always retain their uniform density.
\section{Results}
\label{sec:Results}
A set of simulation runs is performed with different values
for the bending stiffness coefficient $\kappa$. For each simulation,
a normalization run is performed according to the prescription
outlined above. 
The simulation length is 
$10^8$ MC-steps. Figure \ref{plot_Phi}
shows \hbox{$-{\Phi_{\rm RR}}_{,\;ij}(s,\theta)$} for 
\hbox{$\kappa/\kBT=7.5$}.
The rows in the back of the histogram (\hbox{$\theta \rightarrow 0$})
refer to rod pairs in parallel
position; for small $s$, this must be a side-by-side arrangement
(Figure \ref{inclusions} a,b). For \hbox{$s/L \geq 1$},
rod pairs in the in-line formation (Figure \ref{inclusions} d)
also contribute to the average.
The front rows in the diagram (\hbox{$\theta \rightarrow 90^{\circ}$})
refer to rods in the T-formation, which is slightly repulsive
for very small $s$.
This means that a close, perpendicular position 
reduces membrane fluctuations most efficiently.
\Ronfigure{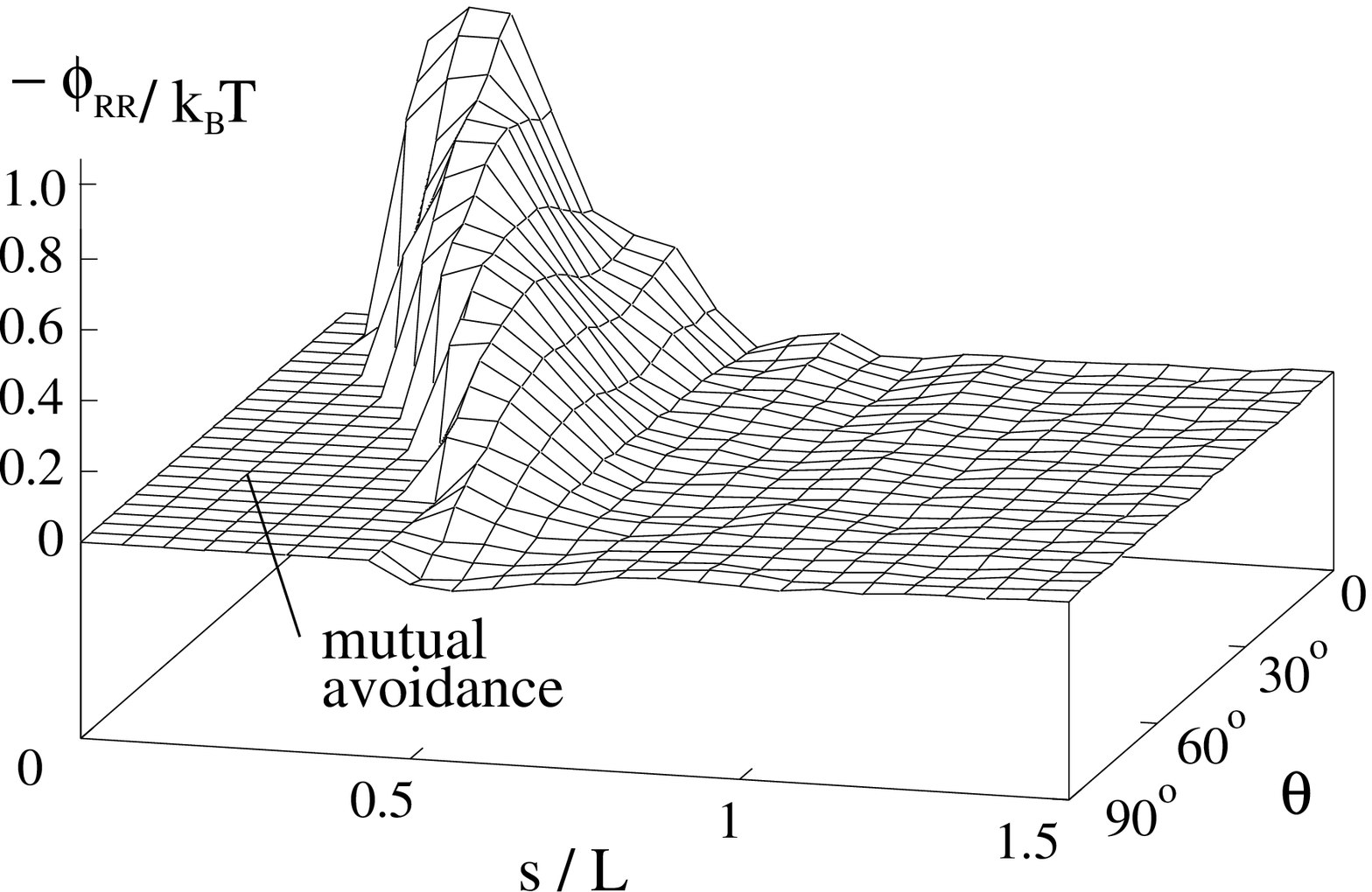}{0.45}{
  Plot of $-\Phi_{\rm RR}(s,\theta)$ for $N=1012$ and 
  \hbox{$\kappa/\kBT=7.5$}.
  The plateau on the left side with the shape of the letter `h'
  is the mutual avoidance area described in 
  Sec.~\ref{sec:Pair distribution functions}; $\Phi_{\rm RR}(s,\theta)$ 
  is not defined there and $\Phi_{\rm ma}$ diverges.
  The peak in the back corresponds to the interactive interaction
  of parallel side-by-side rod pairs, the slight dip in front 
  to rods in the ``T-formation''.
}{plot_Phi}
In order to find out about the impact of many-body effects,
the fraction of entries $n_{ij}$ stemming from isolated pairs
of inclusions to those entries corresponding to clusters of three 
or more inclusions is calculated. An isolated pair is defined as a 
pair of inclusions with \hbox{$s \leq s_{\rm max}$} in the absence of 
other rods which are closer than $s_{\rm max}$ to either of the two.
About $91\%$ of the entries in $n_{ij}$ stem from isolated pairs.
Many-body effects are thus negligible in the present 
simulation as far as the attractive well of the potential is concerned.
These results therefore confirm the conjecture stated at the end of
Sec.~\ref{sec:The_Model}.

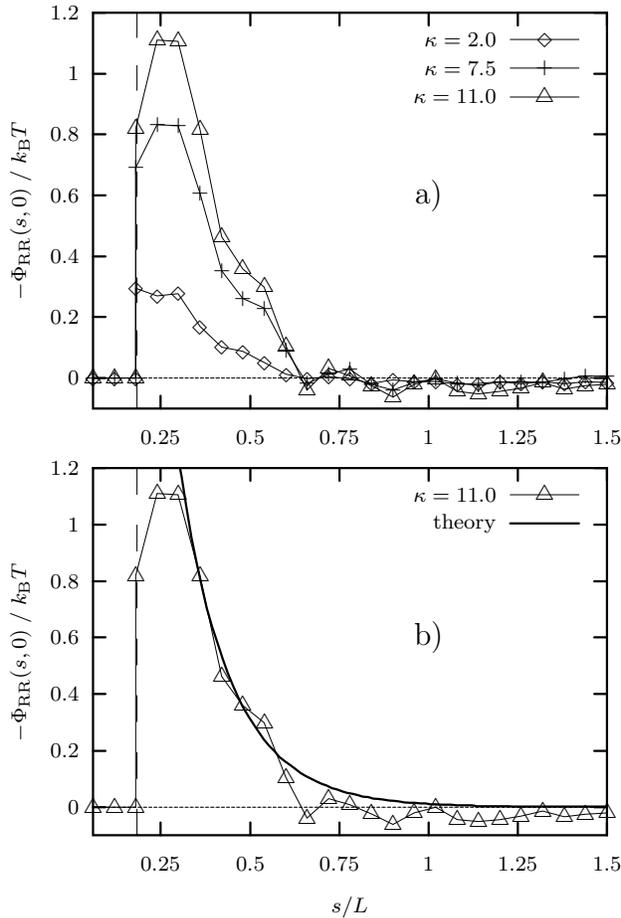
\begin{figure}[p]%
\begin{center}
 \begin{minipage}{\columnwidth}
\setlength{\unitlength}{0.240900pt}
\begin{picture}(1500,900)(0,0)
\footnotesize
\thicklines \path(221,138)(241,138)
\thicklines \path(1028,138)(1008,138)
\put(199,138){\makebox(0,0)[r]{$0$}}
\thicklines \path(221,234)(241,234)
\thicklines \path(1028,234)(1008,234)
\put(199,234){\makebox(0,0)[r]{$0.2$}}
\thicklines \path(221,329)(241,329)
\thicklines \path(1028,329)(1008,329)
\put(199,329){\makebox(0,0)[r]{$0.4$}}
\thicklines \path(221,425)(241,425)
\thicklines \path(1028,425)(1008,425)
\put(199,425){\makebox(0,0)[r]{$0.6$}}
\thicklines \path(221,521)(241,521)
\thicklines \path(1028,521)(1008,521)
\put(199,521){\makebox(0,0)[r]{$0.8$}}
\thicklines \path(221,616)(241,616)
\thicklines \path(1028,616)(1008,616)
\put(199,616){\makebox(0,0)[r]{$1$}}
\thicklines \path(221,712)(241,712)
\thicklines \path(1028,712)(1008,712)
\put(199,712){\makebox(0,0)[r]{$1.2$}}
\thicklines \path(327,90)(327,110)
\thicklines \path(327,712)(327,692)
\put(327,45){\makebox(0,0){0.25}}
\thicklines \path(468,90)(468,110)
\thicklines \path(468,712)(468,692)
\put(468,45){\makebox(0,0){0.5}}
\thicklines \path(608,90)(608,110)
\thicklines \path(608,712)(608,692)
\put(608,45){\makebox(0,0){0.75}}
\thicklines \path(748,90)(748,110)
\thicklines \path(748,712)(748,692)
\put(748,45){\makebox(0,0){1}}
\thicklines \path(888,90)(888,110)
\thicklines \path(888,712)(888,692)
\put(888,45){\makebox(0,0){1.25}}
\thicklines \path(1028,90)(1028,110)
\thicklines \path(1028,712)(1028,692)
\put(1028,45){\makebox(0,0){1.5}}
\thicklines \path(221,90)(1028,90)(1028,712)(221,712)(221,90)
\put(44,401){\makebox(0,0)[l]{\shortstack{\hspace{0.4cm}\rotate{$-\Phi_{\rm RR}(s,0) \; / \; \kBT$}}}}
\put(748,425){\makebox(0,0){\large a)}}
\put(854,670){\makebox(0,0)[r]{$\kappa=2.0$}}
\thinlines \path(876,670)(984,670)
\thinlines \path(221,138)(221,138)(255,138)(288,138)(288,279)(322,267)(355,271)(389,219)(423,187)(456,180)(490,163)(524,144)(557,137)(591,140)(624,137)(658,130)(692,135)(725,132)(759,133)(793,128)(826,130)(860,132)(893,129)(927,132)(961,129)(994,132)(1028,131)
\put(221,138){\raisebox{-1.2pt}{\makebox(0,0){$\Diamond$}}}
\put(255,138){\raisebox{-1.2pt}{\makebox(0,0){$\Diamond$}}}
\put(288,138){\raisebox{-1.2pt}{\makebox(0,0){$\Diamond$}}}
\put(288,279){\raisebox{-1.2pt}{\makebox(0,0){$\Diamond$}}}
\put(322,267){\raisebox{-1.2pt}{\makebox(0,0){$\Diamond$}}}
\put(355,271){\raisebox{-1.2pt}{\makebox(0,0){$\Diamond$}}}
\put(389,219){\raisebox{-1.2pt}{\makebox(0,0){$\Diamond$}}}
\put(423,187){\raisebox{-1.2pt}{\makebox(0,0){$\Diamond$}}}
\put(456,180){\raisebox{-1.2pt}{\makebox(0,0){$\Diamond$}}}
\put(490,163){\raisebox{-1.2pt}{\makebox(0,0){$\Diamond$}}}
\put(524,144){\raisebox{-1.2pt}{\makebox(0,0){$\Diamond$}}}
\put(557,137){\raisebox{-1.2pt}{\makebox(0,0){$\Diamond$}}}
\put(591,140){\raisebox{-1.2pt}{\makebox(0,0){$\Diamond$}}}
\put(624,137){\raisebox{-1.2pt}{\makebox(0,0){$\Diamond$}}}
\put(658,130){\raisebox{-1.2pt}{\makebox(0,0){$\Diamond$}}}
\put(692,135){\raisebox{-1.2pt}{\makebox(0,0){$\Diamond$}}}
\put(725,132){\raisebox{-1.2pt}{\makebox(0,0){$\Diamond$}}}
\put(759,133){\raisebox{-1.2pt}{\makebox(0,0){$\Diamond$}}}
\put(793,128){\raisebox{-1.2pt}{\makebox(0,0){$\Diamond$}}}
\put(826,130){\raisebox{-1.2pt}{\makebox(0,0){$\Diamond$}}}
\put(860,132){\raisebox{-1.2pt}{\makebox(0,0){$\Diamond$}}}
\put(893,129){\raisebox{-1.2pt}{\makebox(0,0){$\Diamond$}}}
\put(927,132){\raisebox{-1.2pt}{\makebox(0,0){$\Diamond$}}}
\put(961,129){\raisebox{-1.2pt}{\makebox(0,0){$\Diamond$}}}
\put(994,132){\raisebox{-1.2pt}{\makebox(0,0){$\Diamond$}}}
\put(1028,131){\raisebox{-1.2pt}{\makebox(0,0){$\Diamond$}}}
\put(930,670){\raisebox{-1.2pt}{\makebox(0,0){$\Diamond$}}}
\put(854,625){\makebox(0,0)[r]{$\kappa = 7.5$}}
\thinlines \path(876,625)(984,625)
\thinlines \path(221,138)(221,138)(255,138)(288,138)(288,469)(322,536)(355,535)(389,428)(423,307)(456,263)(490,248)(524,182)(557,130)(591,144)(624,152)(658,128)(692,119)(725,131)(759,137)(793,130)(826,127)(860,132)(893,132)(927,131)(961,137)(994,141)(1028,141)
\put(221,138){\makebox(0,0){$+$}}
\put(255,138){\makebox(0,0){$+$}}
\put(288,138){\makebox(0,0){$+$}}
\put(288,469){\makebox(0,0){$+$}}
\put(322,536){\makebox(0,0){$+$}}
\put(355,535){\makebox(0,0){$+$}}
\put(389,428){\makebox(0,0){$+$}}
\put(423,307){\makebox(0,0){$+$}}
\put(456,263){\makebox(0,0){$+$}}
\put(490,248){\makebox(0,0){$+$}}
\put(524,182){\makebox(0,0){$+$}}
\put(557,130){\makebox(0,0){$+$}}
\put(591,144){\makebox(0,0){$+$}}
\put(624,152){\makebox(0,0){$+$}}
\put(658,128){\makebox(0,0){$+$}}
\put(692,119){\makebox(0,0){$+$}}
\put(725,131){\makebox(0,0){$+$}}
\put(759,137){\makebox(0,0){$+$}}
\put(793,130){\makebox(0,0){$+$}}
\put(826,127){\makebox(0,0){$+$}}
\put(860,132){\makebox(0,0){$+$}}
\put(893,132){\makebox(0,0){$+$}}
\put(927,131){\makebox(0,0){$+$}}
\put(961,137){\makebox(0,0){$+$}}
\put(994,141){\makebox(0,0){$+$}}
\put(1028,141){\makebox(0,0){$+$}}
\put(930,625){\makebox(0,0){$+$}}
\put(854,580){\makebox(0,0)[r]{$\kappa = 11.0$}}
\thinlines \path(876,580)(984,580)
\thinlines \path(221,138)(221,138)(255,138)(288,138)(288,530)(322,669)(355,667)(389,529)(423,360)(456,310)(490,281)(524,188)(557,119)(591,153)(624,142)(658,126)(692,108)(725,128)(759,137)(793,117)(826,113)(860,117)(893,122)(927,131)(961,121)(994,126)(1028,128)
\put(221,138){\makebox(0,0){$\triangle$}}
\put(255,138){\makebox(0,0){$\triangle$}}
\put(288,138){\makebox(0,0){$\triangle$}}
\put(288,530){\makebox(0,0){$\triangle$}}
\put(322,669){\makebox(0,0){$\triangle$}}
\put(355,667){\makebox(0,0){$\triangle$}}
\put(389,529){\makebox(0,0){$\triangle$}}
\put(423,360){\makebox(0,0){$\triangle$}}
\put(456,310){\makebox(0,0){$\triangle$}}
\put(490,281){\makebox(0,0){$\triangle$}}
\put(524,188){\makebox(0,0){$\triangle$}}
\put(557,119){\makebox(0,0){$\triangle$}}
\put(591,153){\makebox(0,0){$\triangle$}}
\put(624,142){\makebox(0,0){$\triangle$}}
\put(658,126){\makebox(0,0){$\triangle$}}
\put(692,108){\makebox(0,0){$\triangle$}}
\put(725,128){\makebox(0,0){$\triangle$}}
\put(759,137){\makebox(0,0){$\triangle$}}
\put(793,117){\makebox(0,0){$\triangle$}}
\put(826,113){\makebox(0,0){$\triangle$}}
\put(860,117){\makebox(0,0){$\triangle$}}
\put(893,122){\makebox(0,0){$\triangle$}}
\put(927,131){\makebox(0,0){$\triangle$}}
\put(961,121){\makebox(0,0){$\triangle$}}
\put(994,126){\makebox(0,0){$\triangle$}}
\put(1028,128){\makebox(0,0){$\triangle$}}
\put(930,580){\makebox(0,0){$\triangle$}}
\thinlines \drawline[-50](290,90)(290,90)(290,712)
\thinlines \drawline[-50](221,138)(221,138)(229,138)(237,138)(245,138)(254,138)(262,138)(270,138)(278,138)(286,138)(294,138)(303,138)(311,138)(319,138)(327,138)(335,138)(343,138)(351,138)(360,138)(368,138)(376,138)(384,138)(392,138)(400,138)(408,138)(417,138)(425,138)(433,138)(441,138)(449,138)(457,138)(466,138)(474,138)(482,138)(490,138)(498,138)(506,138)(514,138)(523,138)(531,138)(539,138)(547,138)(555,138)(563,138)(572,138)(580,138)(588,138)(596,138)(604,138)(612,138)(620,138)
\thinlines \drawline[-50](620,138)(629,138)(637,138)(645,138)(653,138)(661,138)(669,138)(677,138)(686,138)(694,138)(702,138)(710,138)(718,138)(726,138)(735,138)(743,138)(751,138)(759,138)(767,138)(775,138)(783,138)(792,138)(800,138)(808,138)(816,138)(824,138)(832,138)(841,138)(849,138)(857,138)(865,138)(873,138)(881,138)(889,138)(898,138)(906,138)(914,138)(922,138)(930,138)(938,138)(946,138)(955,138)(963,138)(971,138)(979,138)(987,138)(995,138)(1004,138)(1012,138)(1020,138)(1028,138)
\end{picture}
 \vspace{-2cm}
\setlength{\unitlength}{0.240900pt}
\begin{picture}(1500,900)(0,0)
\footnotesize
\thicklines \path(221,179)(241,179)
\thicklines \path(1028,179)(1008,179)
\put(199,179){\makebox(0,0)[r]{$0$}}
\thicklines \path(221,268)(241,268)
\thicklines \path(1028,268)(1008,268)
\put(199,268){\makebox(0,0)[r]{$0.2$}}
\thicklines \path(221,357)(241,357)
\thicklines \path(1028,357)(1008,357)
\put(199,357){\makebox(0,0)[r]{$0.4$}}
\thicklines \path(221,446)(241,446)
\thicklines \path(1028,446)(1008,446)
\put(199,446){\makebox(0,0)[r]{$0.6$}}
\thicklines \path(221,534)(241,534)
\thicklines \path(1028,534)(1008,534)
\put(199,534){\makebox(0,0)[r]{$0.8$}}
\thicklines \path(221,623)(241,623)
\thicklines \path(1028,623)(1008,623)
\put(199,623){\makebox(0,0)[r]{$1$}}
\thicklines \path(221,712)(241,712)
\thicklines \path(1028,712)(1008,712)
\put(199,712){\makebox(0,0)[r]{$1.2$}}
\thicklines \path(327,135)(327,155)
\thicklines \path(327,712)(327,692)
\put(327,90){\makebox(0,0){0.25}}
\thicklines \path(468,135)(468,155)
\thicklines \path(468,712)(468,692)
\put(468,90){\makebox(0,0){0.5}}
\thicklines \path(608,135)(608,155)
\thicklines \path(608,712)(608,692)
\put(608,90){\makebox(0,0){0.75}}
\thicklines \path(748,135)(748,155)
\thicklines \path(748,712)(748,692)
\put(748,90){\makebox(0,0){1}}
\thicklines \path(888,135)(888,155)
\thicklines \path(888,712)(888,692)
\put(888,90){\makebox(0,0){1.25}}
\thicklines \path(1028,135)(1028,155)
\thicklines \path(1028,712)(1028,692)
\put(1028,90){\makebox(0,0){1.5}}
\thicklines \path(221,135)(1028,135)(1028,712)(221,712)(221,135)
\put(44,423){\makebox(0,0)[l]{\shortstack{\hspace{0.4cm}\rotate{$-\Phi_{\rm RR}(s,0) \; / \; \kBT$}}}}
\put(624,23){\makebox(0,0){$ s/L $}}
\put(748,446){\makebox(0,0){\large b)}}
\put(854,670){\makebox(0,0)[r]{$\kappa = 11.0$}}
\thinlines \path(876,670)(984,670)
\thinlines \path(221,179)(221,179)(255,179)(288,179)(288,543)(322,672)(355,670)(389,542)(423,385)(456,339)(490,312)(524,226)(557,162)(591,193)(624,183)(658,169)(692,152)(725,170)(759,179)(793,160)(826,157)(860,160)(893,165)(927,173)(961,164)(994,168)(1028,170)
\put(221,179){\makebox(0,0){$\triangle$}}
\put(255,179){\makebox(0,0){$\triangle$}}
\put(288,179){\makebox(0,0){$\triangle$}}
\put(288,543){\makebox(0,0){$\triangle$}}
\put(322,672){\makebox(0,0){$\triangle$}}
\put(355,670){\makebox(0,0){$\triangle$}}
\put(389,542){\makebox(0,0){$\triangle$}}
\put(423,385){\makebox(0,0){$\triangle$}}
\put(456,339){\makebox(0,0){$\triangle$}}
\put(490,312){\makebox(0,0){$\triangle$}}
\put(524,226){\makebox(0,0){$\triangle$}}
\put(557,162){\makebox(0,0){$\triangle$}}
\put(591,193){\makebox(0,0){$\triangle$}}
\put(624,183){\makebox(0,0){$\triangle$}}
\put(658,169){\makebox(0,0){$\triangle$}}
\put(692,152){\makebox(0,0){$\triangle$}}
\put(725,170){\makebox(0,0){$\triangle$}}
\put(759,179){\makebox(0,0){$\triangle$}}
\put(793,160){\makebox(0,0){$\triangle$}}
\put(826,157){\makebox(0,0){$\triangle$}}
\put(860,160){\makebox(0,0){$\triangle$}}
\put(893,165){\makebox(0,0){$\triangle$}}
\put(927,173){\makebox(0,0){$\triangle$}}
\put(961,164){\makebox(0,0){$\triangle$}}
\put(994,168){\makebox(0,0){$\triangle$}}
\put(1028,170){\makebox(0,0){$\triangle$}}
\put(930,670){\makebox(0,0){$\triangle$}}
\thinlines \drawline[-50](290,135)(290,135)(290,712)
\thinlines \drawline[-50](221,179)(221,179)(229,179)(237,179)(245,179)(254,179)(262,179)(270,179)(278,179)(286,179)(294,179)(303,179)(311,179)(319,179)(327,179)(335,179)(343,179)(351,179)(360,179)(368,179)(376,179)(384,179)(392,179)(400,179)(408,179)(417,179)(425,179)(433,179)(441,179)(449,179)(457,179)(466,179)(474,179)(482,179)(490,179)(498,179)(506,179)(514,179)(523,179)(531,179)(539,179)(547,179)(555,179)(563,179)(572,179)(580,179)(588,179)(596,179)(604,179)(612,179)(620,179)
\thinlines \drawline[-50](620,179)(629,179)(637,179)(645,179)(653,179)(661,179)(669,179)(677,179)(686,179)(694,179)(702,179)(710,179)(718,179)(726,179)(735,179)(743,179)(751,179)(759,179)(767,179)(775,179)(783,179)(792,179)(800,179)(808,179)(816,179)(824,179)(832,179)(841,179)(849,179)(857,179)(865,179)(873,179)(881,179)(889,179)(898,179)(906,179)(914,179)(922,179)(930,179)(938,179)(946,179)(955,179)(963,179)(971,179)(979,179)(987,179)(995,179)(1004,179)(1012,179)(1020,179)(1028,179)
\put(854,625){\makebox(0,0)[r]{theory}}
\thicklines \path(876,625)(984,625)
\thicklines \path(358,712)(360,699)(368,648)(376,603)(384,562)(392,525)(400,492)(408,463)(417,436)(425,411)(433,390)(441,370)(449,352)(457,336)(466,321)(474,308)(482,296)(490,285)(498,275)(506,267)(514,258)(523,251)(531,245)(539,239)(547,233)(555,228)(563,224)(572,220)(580,216)(588,213)(596,209)(604,207)(612,204)(620,202)(629,200)(637,198)(645,196)(653,195)(661,193)(669,192)(677,191)(686,190)(694,189)(702,188)(710,187)(718,186)(726,186)(735,185)(743,185)(751,184)(759,184)
\thicklines \path(759,184)(767,183)(775,183)(783,183)(792,182)(800,182)(808,182)(816,182)(824,181)(832,181)(841,181)(849,181)(857,181)(865,181)(873,181)(881,180)(889,180)(898,180)(906,180)(914,180)(922,180)(930,180)(938,180)(946,180)(955,180)(963,180)(971,180)(979,180)(987,180)(995,180)(1004,180)(1012,180)(1020,180)(1028,180)
\end{picture}
 \caption{  {
  a) Plot of \hbox{$-\Phi_{\rm RR}(s,\theta=0)$};
  the graph for $\kappa/\kBT = 7.5$ ($+$) corresponds to the 
  slice from Figure \ref{plot_Phi} along $s$ at $\theta=0$.
  The distance between two tics on the abscissa equals one 
  mean bond length $\avg{r}$ ($L=4\avg{r}$). The vertical dashed line
  at $s/L = a/L = 0.183$ marks the smallest possible bond length.
  b) as a) but for \hbox{$\kappa/\kBT=11.0$} together with a fit of
  \ \hbox{$a_1 \ln(1-\exp(-a_2 s/L))$}, ($a_1=7.5$, $a_2=6.6$) 
  following the theory of Netz [7]. 
} \label{plot_rod_slice} }
 \end{minipage}
\end{center}
\end{figure}

In Figure \ref{plot_rod_slice}a, \hbox{$-\Phi_{\rm RR}(s, \theta=0)$}
is shown for $\kappa/\kBT = 2.0,\; 7.5$, and $11.0$. The decay
is very rapid with $\Phi_{\rm RR}$ vanishing approximately at
$s/L \approx 0.6$ within the precision of the simulation. 
At $s/L = 0.75 = 3\avg{r}$ and $s/L = 1.0 = 4\avg{r}$,
remnants of the lattice periodicity induce small peaks in the graphs.
The figure shows that $\Phi_{\rm RR}$ is attractive for small 
distances $s \stackrel{~}{<} L$, but short-range.
Netz \cite{Net97b} found \hbox{$\Phi_{\rm RR}(s)$} to decay logarithmically.
A fit with a logarithmic function of the type
given in \cite{Net97b} is shown in Figure \ref{plot_rod_slice}b
\footnote{This refers to inclusions with a quadratic term
in the perturbation Hamiltonian \cite{Net97b}.}.
In addition to the short-range attraction there might also be a long-range
component. In \cite{Gol96b} $\kBT/128$ was estimated for the magnitude of the
leading term of long-range attraction in the effective potential
between rodlike inclusions. This value is less than the statistical error
of our data in the limit of larger s (see Figure \ref{plot_rod_slice}). 

Consequently, our data
do not permit us to comment on long-range effective interactions between
inclusions. It is, however, noteworthy, that
due to the computational method of a {\em perturbative} treatment
in terms of inverse powers of $s$, the results of 
\cite{Gou93,Gol96b} cannot rule out the existence of short-range 
attractions consistent with the present findings. The results in
\cite{Net97b}, on the other hand, are {\em exact} and remain
valid in the strong-coupling limit.

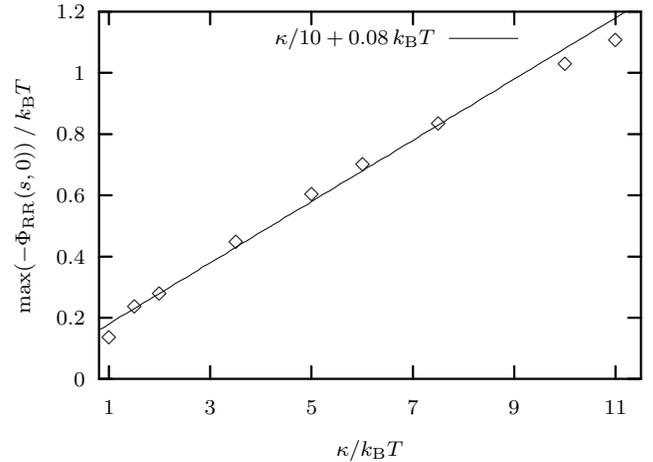
\begin{figure}[tbp]%
  \center{  
\setlength{\unitlength}{0.240900pt}
\begin{picture}(1500,900)(0,0)
\footnotesize
\thicklines \path(177,135)(197,135)
\thicklines \path(1028,135)(1008,135)
\put(155,135){\makebox(0,0)[r]{0}}
\thicklines \path(177,231)(197,231)
\thicklines \path(1028,231)(1008,231)
\put(155,231){\makebox(0,0)[r]{0.2}}
\thicklines \path(177,327)(197,327)
\thicklines \path(1028,327)(1008,327)
\put(155,327){\makebox(0,0)[r]{0.4}}
\thicklines \path(177,424)(197,424)
\thicklines \path(1028,424)(1008,424)
\put(155,424){\makebox(0,0)[r]{0.6}}
\thicklines \path(177,520)(197,520)
\thicklines \path(1028,520)(1008,520)
\put(155,520){\makebox(0,0)[r]{0.8}}
\thicklines \path(177,616)(197,616)
\thicklines \path(1028,616)(1008,616)
\put(155,616){\makebox(0,0)[r]{1}}
\thicklines \path(177,712)(197,712)
\thicklines \path(1028,712)(1008,712)
\put(155,712){\makebox(0,0)[r]{1.2}}
\thicklines \path(193,135)(193,155)
\thicklines \path(193,712)(193,692)
\put(193,90){\makebox(0,0){1}}
\thicklines \path(352,135)(352,155)
\thicklines \path(352,712)(352,692)
\put(352,90){\makebox(0,0){3}}
\thicklines \path(511,135)(511,155)
\thicklines \path(511,712)(511,692)
\put(511,90){\makebox(0,0){5}}
\thicklines \path(670,135)(670,155)
\thicklines \path(670,712)(670,692)
\put(670,90){\makebox(0,0){7}}
\thicklines \path(829,135)(829,155)
\thicklines \path(829,712)(829,692)
\put(829,90){\makebox(0,0){9}}
\thicklines \path(988,135)(988,155)
\thicklines \path(988,712)(988,692)
\put(988,90){\makebox(0,0){11}}
\thicklines \path(177,135)(1028,135)(1028,712)(177,712)(177,135)
\put(44,423){\makebox(0,0)[l]{\shortstack{\rotate{${\rm max}(-\Phi_{\rm RR}(s,0)) \:/\: \kBT$}}}}
\put(602,23){\makebox(0,0){$\kappa/\kBT$}}
\put(193,202){\raisebox{-1.2pt}{\makebox(0,0){$\Diamond$}}}
\put(233,250){\raisebox{-1.2pt}{\makebox(0,0){$\Diamond$}}}
\put(272,270){\raisebox{-1.2pt}{\makebox(0,0){$\Diamond$}}}
\put(392,351){\raisebox{-1.2pt}{\makebox(0,0){$\Diamond$}}}
\put(511,426){\raisebox{-1.2pt}{\makebox(0,0){$\Diamond$}}}
\put(591,473){\raisebox{-1.2pt}{\makebox(0,0){$\Diamond$}}}
\put(710,537){\raisebox{-1.2pt}{\makebox(0,0){$\Diamond$}}}
\put(909,631){\raisebox{-1.2pt}{\makebox(0,0){$\Diamond$}}}
\put(988,668){\raisebox{-1.2pt}{\makebox(0,0){$\Diamond$}}}
\put(705,670){\makebox(0,0)[r]{$\kappa/10+0.08\, \kBT$}}
\thinlines \path(727,670)(835,670)
\thinlines \path(177,212)(177,212)(186,217)(194,222)(203,228)(211,233)(220,238)(229,243)(237,248)(246,254)(254,259)(263,264)(272,269)(280,274)(289,279)(297,285)(306,290)(315,295)(323,300)(332,305)(340,311)(349,316)(358,321)(366,326)(375,331)(383,337)(392,342)(400,347)(409,352)(418,357)(426,363)(435,368)(443,373)(452,378)(461,383)(469,389)(478,394)(486,399)(495,404)(504,409)(512,415)(521,420)(529,425)(538,430)(547,435)(555,441)(564,446)(572,451)(581,456)(590,461)(598,467)
\thinlines \path(598,467)(607,472)(615,477)(624,482)(633,487)(641,493)(650,498)(658,503)(667,508)(676,513)(684,519)(693,524)(701,529)(710,534)(719,539)(727,545)(736,550)(744,555)(753,560)(762,565)(770,571)(779,576)(787,581)(796,586)(805,591)(813,597)(822,602)(830,607)(839,612)(847,617)(856,622)(865,628)(873,633)(882,638)(890,643)(899,648)(908,654)(916,659)(925,664)(933,669)(942,674)(951,680)(959,685)(968,690)(976,695)(985,700)(994,706)(1002,711)(1004,712)
\end{picture}
\caption{
  Plot of \hbox{${\rm max}(-\Phi_{\rm RR}(s,\theta=0))$} for different 
  bending moduli $\kappa$.
  Inclusions on stiffer membranes feel a more attractive potential,
  the relationship is roughly linear with 
  \hbox{${\rm max}(-\Phi_{\rm RR}(s,0)) \approx \kappa/10$}.
} \label{plot_rod_depth} }
\end{figure}

The larger $\kappa$, the
more attractive becomes the interaction. 
Figure \ref{plot_rod_depth}
shows that this relationship is almost linear, with roughly
\hbox{${\rm max}(-\Phi_{\rm RR}(s,\theta=0)) \approx \kappa/10$}.

In other words, the effective
attractive interaction between a pair of inclusions becomes stronger the
stiffer the membrane of the vesicle is. This may seem contradictory
because the stiffer the membrane the more suppressed are fluctuations of
the membrane. 
However, we notice from the work of Netz \cite{Net97b} that
short-range interactions between a pair of inclusions may be expected to 
depend logarithmically on the ratio $s/\xi$ for $s/\xi \ll 1$. 
A fit of the functional form for
$\Phi_{\rm RR}(s)$ proposed by Netz to our data shows that the latter are 
consistent with a logarithmic decay of the effective potential 
(see Figure \ref{plot_rod_slice}b). The
correlation length $\xi$, on the other hand, increases with $\kappa$ as
\hbox{$\xi \propto \exp(4 \pi \kappa/(3 \kBT))$} \cite{Ips95}. 
Thus, combining the logarithmic decay of
$\Phi_{\rm RR}$ with the latter expressions leads to a linear increase of 
$-\Phi_{\rm RR}$ with $\kappa$ for fixed $s$ and $T$ which is
consistent with the plot in Figure \ref{plot_rod_depth}.
\section{Discussion and Conclusions}
\label{sec:Discussion and Conclusions}
The present article discusses Monte Carlo simulations for closed,
triangulated membranes with mobile, 
rod-like inclusions as an extension of a widely used 
model \cite{Gom97a,Gom96,Ips95,Kan87a}.
Inclusions locally straighten the network, 
thereby quenching lateral fluctuations.
The effective interaction potential is attractive and short-range
(see Figure \ref{plot_rod_slice}). For $s \rightarrow 0$, it is
limited by the mutual avoidance condition (Figures \ref{winkel}
and \ref{plot_Phi}).

The contribution of this simulation lies in the focus on small
inclusion separations. 
We find an attractive interaction potential of the 
order of $\kBT$ between rods that consist of five rigidly 
connected hard spheres, for bending coefficients of 
$\kappa/\kBT$ of the order of
$1-10$ (Figure \ref{plot_rod_slice}). 
Furthermore, the magnitude of the 
well depth of the potential grows almost linearly with $\kappa$,
as shown in Figure \ref{plot_rod_depth}.
All of the results presented in this paper
are valid in the dilute inclusion limit.
The form of the function $\Phi_{\rm RR}$ and its magnitude
can be expected to change with higher densities. For very high
inclusion densities, segregated phases might exist, one of them 
largely depleted of inclusions and another, dense phase with a nematic 
order due to the inclusion anisotropy.

In conclusion, these results have implications for the formation
of inclusion clusters. Such aggregation is frequently observed
experimentally \cite{Ste87} and in simulation \cite{Sab98}. 
A number of different driving 
mechanisms have been proposed, aside from those induced by 
direct interaction or depletion forces \cite{Yam98a,Goem98}.
This includes spontaneous curvature \cite{Net95a},
lateral tension \cite{Fou96,Wei98}, and
conical inclusion shapes \cite{Fou96,Wei98,Dom98}.
Fluctuation-induced interactions as considered in 
\cite{Gou93,Par96,Gol96b} focus on the large-separation limit
$s/L \gg 1$ and result in interaction potentials 
$\Phi_{\rm RR} \ll \kBT$. Although this is of great 
theoretical interest as it draws an analogy to the 
quantum-mechanical Casimir 
effect \cite{Cas48,KRE94}, interaction energies far smaller than 
$\kBT$ are of little practical relevance in thermodynamic systems.
The present work, in contrast, allows to gain insight into the strongly
perturbed, short length-scale region, over distances comparable to both
the inclusion length and the lipid head size. This is the crucial length
scale in chemical and biological applications.
\section*{Acknowledgements}
\label{sec:Acknowlegdements}
RH would like to thank G.~Gompper (Max\--Planck\--Institut
f\"ur Kolloid- und Grenz\-fl\"achen\-for\-schung Teltow, Germany) and
M.~Kr\"oger for many stimulating and helpful discussions.
MS is grateful to the Son\-der\-for\-schungs\-bereich 448
``Me\-so\-sko\-pisch struk\-tu\-rierte Ver\-bund\-sy\-steme'' for 
financial support.
\newcommand{\BIBpra}{Phys. Rev. A\ }
\newcommand{\BIBprb}{Phys. Rev. B\ }
\newcommand{\BIBpre}{Phys. Rev. E\ }
\newcommand{\BIBprl}{Phys. Rev. Lett.\ }
\newcommand{\BIBeuro}{Europhys. Lett.\ }
\newcommand{\jpIF}{J. de Phys. I France\ }
\newcommand{\jpIIF}{J. de Phys. II France\ }
%

%
%
%
%
\end{multicols}
\end{document}